*Article*

# Implementing biosensing-based user preference visualisation in architectural spaces

Mi Kyoung Kim

School of Architecture, Hanyang University, Seoul 04763, Republic of Korea; musicji83@hanyang.ac.kr (M.K.K.);
Tel.: +82-2-2220-1796

**Abstract:** This study delves into the interplay between architectural spaces and human emotions, leveraging the emergent field of neuroarchitecture. It examines the functional and aesthetic influence of architectural design on individual users, with a focus on biosensing data such as brainwave and eye-tracking information to understand user preferences. Highlighting the case of the unbuilt Dongdaemun Design Plaza by Zaha Hadid, the study reflects on the divided opinions regarding architectural success and its emotional impact on the public. It features the work of neuroscientist Oshin Vartanian, who has found that the form of architectural spaces—curved versus straight—affects brain activity patterns, specifically in the orbitofrontal and entorhinal cortices. The advancement in portable EEG equipment and AI-driven monitoring software has broadened the scope of neuroarchitectural research, enabling a more nuanced analysis of emotional responses within architectural spaces. The study aims to visualize these emotional responses through heat maps in virtual reality environments, offering a novel method to capture and express users' preferences and emotional states in relation to architectural designs. This approach suggests a future where architectural spaces are optimized for emotional engagement, informed by direct neural and visual feedback from the intended users.

## 1. Introduction

*1.1. Background*

The space created by architecture is a functional and aesthetic environment that directly or indirectly affects human emotions, and the process of creating an optimal design through the architect's design research is carried out. The proposed space plays a functional and aesthetic role as a space for individual users. These architectural spaces are used as a factor that affects the emotions of the users of the space, and various attempts are made to satisfy both the form and function of the building. In 2013, a DDP designed by Zaha Hadid in Seoul was left unbuilt, sparking controversy. While some criticised it as a landmark building that did not reflect the coldness of the metal exterior panels and the locality of the site, others praised it for creating a public building for citizens through cutting-edge architectural design in an area that had declined after industrialisation, and most experts criticised it as a strategic building for Seoul, which was trying to become a world-class attraction through a renowned architect. In this study, however, we aimed to analyse these controversial buildings from the perspective of space users, using biosensing data to find out which parts of the space are highly preferred for information about future buildings. Biosensing data refers to data extracted from sensors to monitor people, and in this study only brainwave and eye-tracking information was used.

Neuroscientist Oshin Vartanian, who has been studying the social phenomenon of architectural space by focusing on the relationship between the form of architectural space and people, has tested the claim that the pattern of brain waves can change depending on the form of architectural space. In particular, he found that although different users have different preferences, curved versus straight buildings produced distinct differences in the magnitude of change in EEG activity patterns in the orbitofrontal cortex and entorhinal cortex, two parts of the human brain. Studies such as these are leading to the development of neuroarchitecture, a field of study that analyses the psychological state of users in architectural spaces. Neuroarchitecture is a study that seeks to analyse the emotional information of users in architectural spaces by analysing EEG patterns, based on themes that have been explored in the field of environmental psychology. Recently, as EEG equipment has become more portable and monitoring software based on artificial intelligence has advanced, the scope of research has expanded as the constraints of EEG experiments in architectural spaces have been reduced.

In recent years, devices that monitor human biometric information have evolved from simple medical purposes as sensors have become more sophisticated and miniaturised. In various design fields, including architecture, where users' design preferences are important, researchers have used EEG and eye-tracking information to predict consumers' preferences for design alternatives in order to determine the final design. By using biometric information that combines EEG and eye-tracking information, this study aims to visually express the emotional information that users feel in architectural spaces using heat maps. Although individual users' preferences in architectural spaces may vary, positive and negative preferences can be identified through heat maps and can help to understand the relationship between a place and its design for the type of people who use it. By analysing the user's cognitive process of design forms in architectural spaces and visualising the emotional information felt by the design in the space through a three-dimensional heat map, it will be easy to visually perceive the emotional information felt by the user in the space.

John P. Eberhard has argued that EEG research will lead to a better understanding of how architecture can change the way we design buildings, postulating that a better understanding of how buildings and places affect the mental states of their users will lead to a better understanding of how the human brain works, and has conducted research that describes architectural design criteria for schools, offices, laboratories, memorials, churches and facilities for the elderly, and offers hypotheses about the human experience in such environments.

John Gero hypothesised that open spaces are more effective for creativity than closed spaces and conducted a study to compare the EEG analysis of 18 professional industrial designers to find the correlation between the openness of space and brain waves using protocol analysis techniques. Many experts and scholars have conducted studies to monitor users' EEG in architectural environments, analyse the correlation between space and users, and verify hypotheses, but there is a lack of research to develop monitoring analysis and visualisation methods suitable for the purpose of experiments on architectural spaces using EEG data and eye tracking information. In particular, we did not find any studies that attempted to visualise emotional information as a heat map using EEG data and eye-tracking information collected while watching 360-degree videos in VR environments, as is done in this study.

Therefore, this study aims to present a method to express the preferred areas in the architectural space as a heat map by integrating EEG information and eye tracking information generated in the body into a heat map in the architectural space environment used by users. In particular, as a difference between the existing research and this research, we aim to implement the architectural space in VR space as a 360-degree image without an experimenter visiting a specific place, and collect EEG information and eye tracking information in the VR environment to implement emotional information about the user's architectural space through a heat map, and use biosensing information as a core technology to reflect the information experienced in the virtual space through heat map technology in VR space by implementing the site of the real space in VR360.

*1.2. Methodology*

The existing heat map implementation method using eye-tracking equipment has the limitation that the heat map of the site can only be performed through a limited reference image in the space as shown in Figure 1. Therefore, this study aims to overcome this limitation and build a heat map in three dimensions by mapping 360-degree images to a VR environment map.

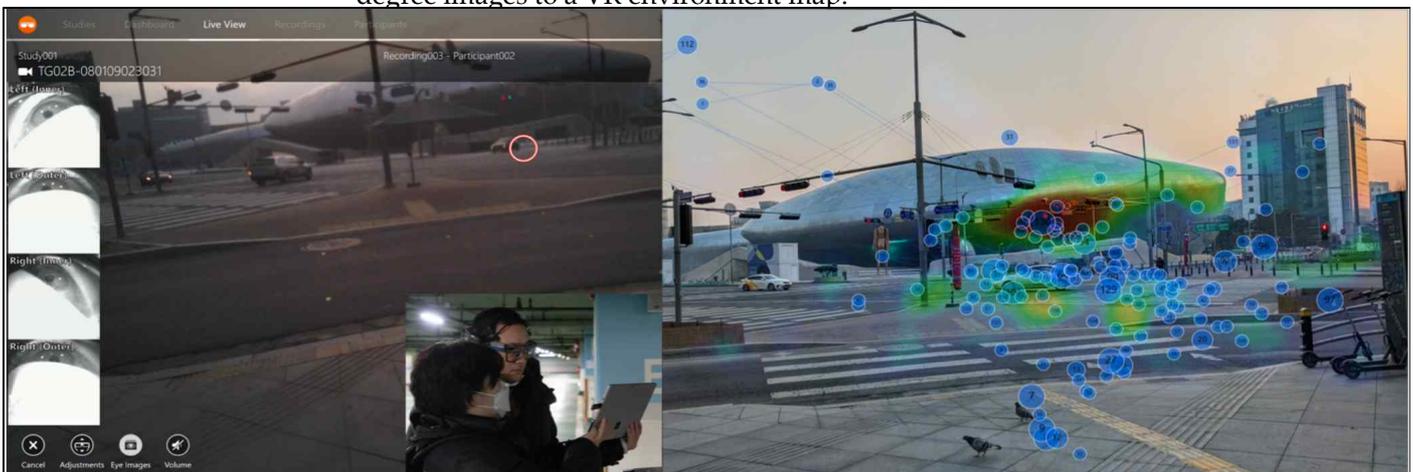

[Figure 1: Traditional eye tracking heatmap visualisation

This study explores the core technology of implementing indicators of emotional information through 360-degree heat maps using EEG and eye-tracking information as a means of identifying users' preferences for architectural spaces by experiencing 360-degree videos of the experimental space in a VR environment without visiting the site. In order to reflect the information of residents' preferences for architectural spaces through the use of emotional map visualisation, this study partially applies the method of 'using maps to collect emotional information' among the three categories of emotional maps proposed by Griffin & Mcquoid (2012). In Griffin's case, emotional information was reflected in a two-dimensional map environment, but in this study, emotional information is reflected in a three-dimensional environment map of the VR environment, so the method of mapping emotional information by changing from two to three dimensions is different, but the method of mapping by quantifying brain wave information is similar. Through this method, it can be measured and visualised as auxiliary data to provide optimised architectural space information for residents in the future architectural planning stage, and architects can refer to it when designing architectural spaces. The methodological process for implementing the research involves building a 360-degree video image of the target architectural space in a VR environment in a game engine and building a model that fuses EEG and eye tracking information based on deep learning algorithms. The collected information is used to propose a process in which an artificial intelligence algorithm derives fixed coordinates in the VR environment based on the visual preferences of long-lasting gaze and emotional information of EEG. The scope of this study is to construct a heat map system using two types of biometric data, and to present fixed environment map coordinates for 360 videos in a VR environment by analysing the EEG sensor data of a user with low stress and high eye tracking information with an artificial intelligence model. In particular, we focused on the process of linking EEG data, eye-tracking information, and VR environment maps, and conducted experiments on a single experimenter to ensure that the heat map visualisation task proceeds smoothly (Figure 2).

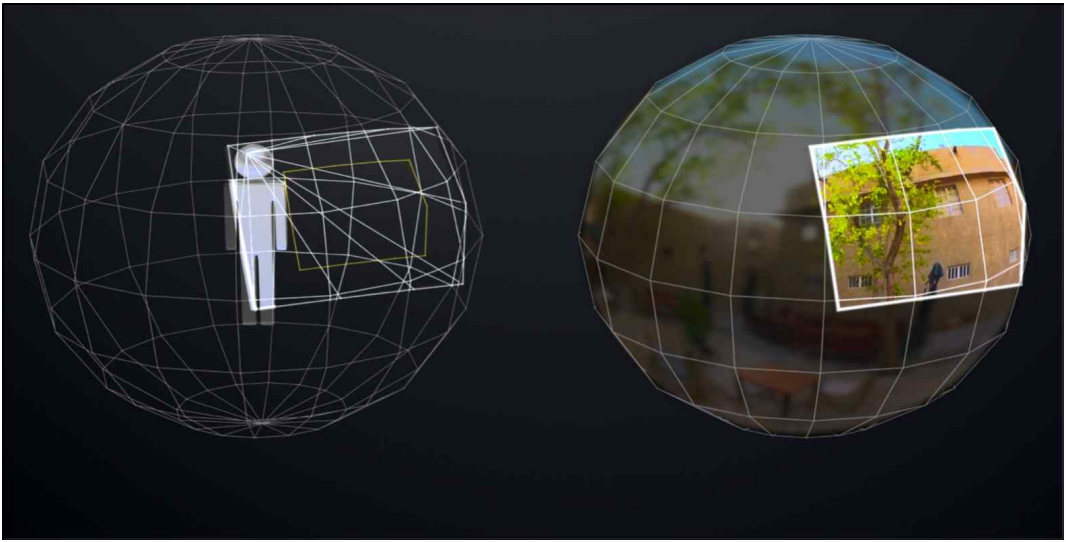

[Figure 2] Implementation of eye tracking heatmap visualisation using VR environment maps

## 2. Mechanisms of fusion of EEG and eye tracking in architectural spaces

The study of the correlation between architectural space and human emotions has been studied under the name of environmental psychology, and the study of how architectural space affects the human mind has evolved into the study of neuroarchitecture through psychology and neuroscience that explores the human mind. Neuroarchitecture seeks to analyse human psychological changes in response to architectural spaces by measuring brain activity to test the hypotheses defined by environmental psychology. In the last decade, EEG equipment has been miniaturised through the development of sensors, and attempts have been made to use it in various fields, mainly for medical purposes, creating conditions for easy collection and monitoring of EEG information without complex preparation. This has created a pipeline for measuring the impact on human emotions in architectural spaces through neuroarchitecture. In this study, EEG was utilised to monitor user emotional information. EEG is a time series of microcurrent data generated by the brain, and various artificial intelligence algorithms have been used to derive emotional information from it. EEG-based brainwave data is classified into Alpha, Beta, Gamma, Delta, and Theta through a power spectrum-based refinement process from raw data, and in this study, six emotional index data were measured through Emotiv's API, Performance Metric algorithm based on SVM vector machine. This API has been continuously learned and developed since 2012, and the reliability verification of 'EEG Based Emotion Recognition' has been covered in various articles through experiments and related papers. The emotional information that Performance Metric classifies from EEG information is shown in Table 1.

Table 1 Taxonomy of sentiment by performance metrics

| Data type | EEG data characteristics |
|---|---|
| Stress | Tension indicates the degree of tension for the current stimulus. High tension tends to be destructive and can have a negative impact on health. |
| Engagement | Immersion refers to the level of engagement in a moment and refers to the conscious direction of arousal and attention to a stimulus. Engagement is also characterised by a mixture of attention and focus. |
| Interest | Affinities represent preferences and dislikes for environments and stimuli. A high affinity indicates a strong liking, while a low affinity indicates a strong dislike. |
| Excitement | Arousal refers to the perception and feeling of physiological arousal. A high arousal level indicates high physiological arousal. |
| Focus | Concentration is the degree of fixed attention to a particular task based on the depth of interest and the frequency of switching attention between tasks. A lower concentration number indicates higher interest. |
| Relaxation | Relaxation is the ability to recover from a state of intense concentration. Higher levels of relaxation indicate a higher capacity for rest and recovery. |

Eye tracking research is a study of visual attention to stimuli through analysis of gaze fixation and gaze movement according to gaze direction for human interests and preferences in architectural spaces. Recently, studies have been conducted to combine eye tracking information with EEG signals to analyse users' design preferences. An example of a study combining EEG and eye tracking to analyse human physiological responses is a marketing study for Macromill's Asahi Mogitate beverage pack design. The study analysed consumer preferences for the design of Asahi's canned beverages, and the final design was selected by analysing brain waves based on consumers' interest in and positivity towards the can design alternatives and the fixation time of their gaze on the text information on the can. This combination of EEG and eye-tracking information was used to index preferences and visualise preference information based on biosensory information through heat maps. In the case of cans and other similarly sized products, eye-tracking devices are often used in product design because they are within the range of the camera's field of view (FOV). However, there are some cases, such as architectural spaces, that cannot be covered by the camera's FOV at once, so this study attempts to overcome the limitations of the FOV of the eye tracking camera through a VR environment (Figure 3).

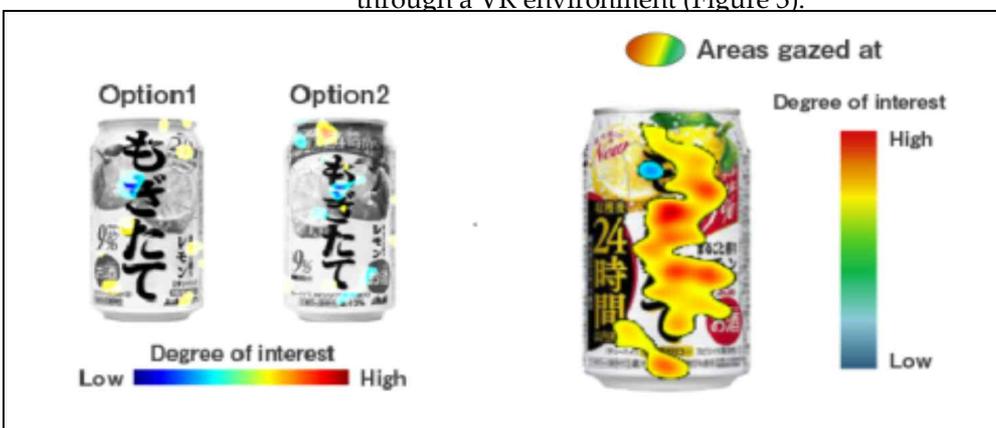

[Figure 3: Combining EEG and eye tracking to develop a new beverage pack design(Source: Macromill Group, 2019)

An example of a study using EEG and eye-tracking for architectural spaces is Choi & Kim (2019), who monitored EEG and eye-tracking data through images of cases that changed the physical information of architectural spaces and analysed the differences in preferences for each alternative. Through these examples of previous studies combining EEG and eye tracking, we have analysed the emotional data felt by users, derived an index of interest or stress in a space, and identified visualisation methods using heat maps. However, in the case of previous studies, users' preferences were analysed through the monitoring process of brainwave information and eye-tracking data that users felt superficially without being immersed in the scene through images organised into two-dimensional images and monitors (Figure 4).

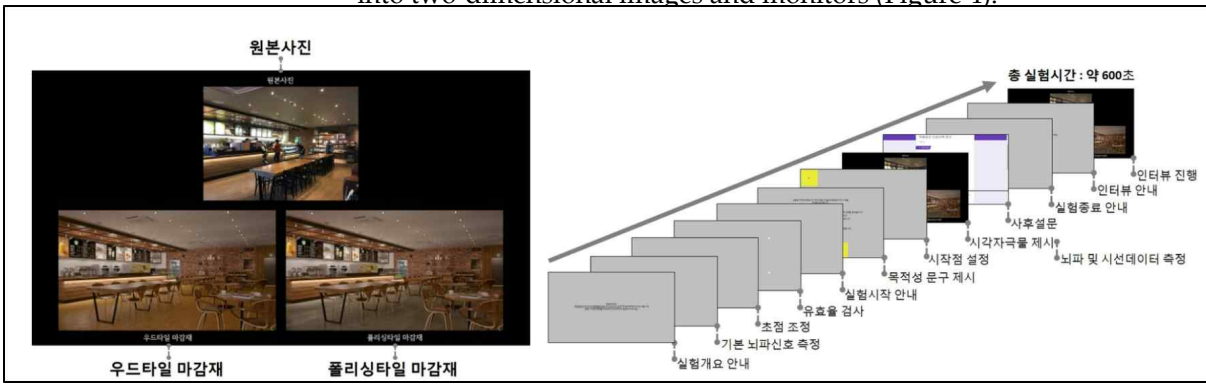

[Figure 4] Architecture study combines EEG and eye tracking(Source: Choi & Kim, 2019)

In addition, existing analytics platformsthat integrate EEG and eye-tracking information are measuring preferences for design alternatives represented on objects such as cans, or analysing emotional information from rendered images by restricting the observer's view to only a small portion of the architectural space, as in the prior art. We believe that this is due to the limitations of currently available applications that integrate and analyse EEG and eye-tracking data, as well as the low demand for analysing immersive environments such as architectural spaces. Therefore, in this study, we built an immersive experimental environment with a VR environment for architectural spaces and implemented a three-dimensional heat map instead of a two-dimensional heat map using EEG and eye tracking data. In the case of conventional heat map visualisation, architectural spaces are identified through images with a fixed composition and biometric data is reflected in the heat map, but in the case of switching to a VR environment, 360-degree images are mapped on a three-dimensional spherical environment map to increase the immersion of the experimenter. In addition, we aimed to improve the limitations of limited emotional information visualisation by implementing a heat map in the entire three-dimensional spherical shape instead of implementing a heat map limited to two-dimensional images.

In the architectural space, the user's biosensing information, brain wave information and gaze estimation information, accumulate 60 and 50 data per second at 60Hz and 50Hz, respectively. In a typical experiment of only 10 minutes, more than 30,000 data were accumulated, and it took a lot of time to process and analyse the data through the change value of the relationship between the gaze duration and the index of interest and stress according to the EEG, and human error occurred during manual data processing. To solve these problems, we used the 'Multivariable Linear Regression' algorithm to analyse human preferences for architectural spaces using a deep learning-based ranking system among artificial intelligence algorithms. The algorithm analyses the input data and lists the rankings using human preferences and gaze delay information. Therefore, Multivariable Linear Regression-based ranking algorithm is applied to the emotional indicator data of Performance Metrics measured through experiments with time information as the axis. By using a proven deep learning algorithm for the purpose of this study, we aimed to automate the existing data analysis method and increase the reliability of the data.

The deep learning ranking system for biosensing-based preferences in this study involved a three-step process: (Figure 6)

First, the emotional indexes Interest and Stress from the EEG data and the coordinate values of the eye tracking information and the delay values of the gaze were placed on the same time series axis.

Second, one of the algorithms of the deep learning ranking system, "Multivariable Linear Regression", is used to derive the highest priority ranking for areas of high interest and low stress, as well as areas where the gaze lingers for a long time.

Third, derive the top 30 preference data from the deep learning ranking system and the coordinate data for the heatmap and visualise it as a heatmap.

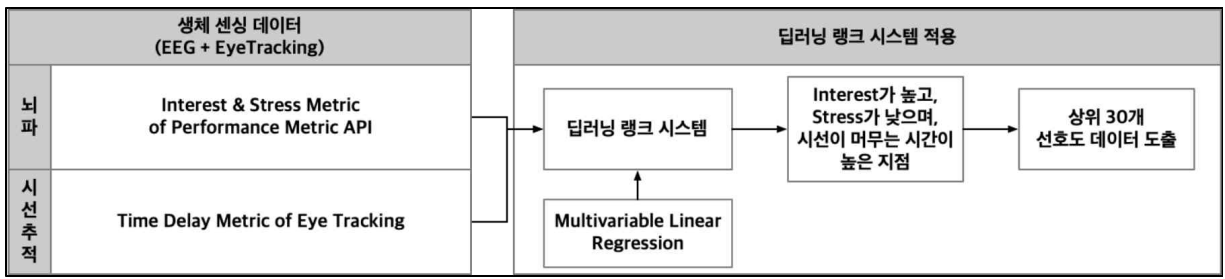

[Figure 6] Deep learning ranking system algorithm for biosensing-based preferences

### 3. Build a user-preference biosensing-based VR heatmap process using EEG and eye-tracking information in architectural spaces

The biosensing-based VR heatmap process proposed in this study consists of three steps.

First, collecting user's biometric data throughVR environment of brainwave data and eye tracking data. The heat map of the existing eye tracking equipment has the limitation that the visualisation is performed through the heat map only for the specified reference image. Therefore, eye tracking sensors are mounted on VR equipment to collect human eye tracking information, and EEG equipment in the form of a bandana is worn to collect user emotional information. This method has the advantage of recording and expressing heat maps in the entire 360-degree image, beyond the limitation of collecting limited heat maps through designated images. In this study, the DDP area was selected for biometric data measurement. Three locations of DDP architectural space were selected and a total of 10 minutes of 360-degree video footage of the architectural space to be reflected in the 3D environment map of VR was taken. The reason for using 360-degree video rather than the usual 3D modelling method for building VR environments was to enhance immersion through the sounds and images of real-world environments. The captured video footage is used to create a virtual reality space through VR equipment capable of eye tracking, and measurements are performed through eye tracking information and EEG measurement equipment.

Second, in the biometric data refinement task, each raw data is transformed using the API for preference visualisation. Among the user's biometric data collected through the VR environment, the brainwave data uses the Performance Metrics API to organise the user's emotional information into six index systems. In the case of eye tracking information, the eye tracking sensor of HTC VIVE Pro Eye was used as a sensor to track the position of the pupil, and an API was developed to organise the coordinate data of the point where the user's gaze intersects with the coordinate system of the 360-degree video played on the VR environment map. (Figure 7: Diagram representation of the coordinate system of the 360-degree video and the intersection of the gaze (pupil) to calculate the coordinates.)

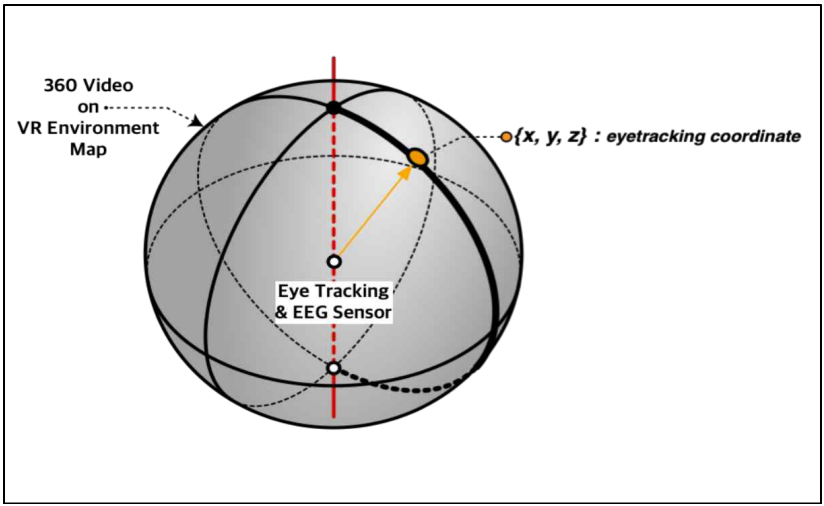

[Figure 7] Diagram of 360 video coordinate system with intersecting gaze to calculate coordinates (future)

Third, the VR heatmap preference visualisation uses a deep learning-based ranking system to list EEG data and eye tracking information according to preference on the same time series axis, and uses Unity3D to display the preferences of the space experiencer's preferred areas of the architectural space as a heatmap in a three-dimensional virtual reality space. The main role of the VR environment map in the form of a globe is to derive the coordinates that have a high ranking of positive emotional states through brain waves and a high ranking of long gaze duration using a deep learning-based ranking algorithm and express them as a heat map. In the case of a three-dimensional heat map through VR, there is a limitation that the heat map can only be checked in the VR environment, so we want to visualise and express it as a Mercator type by spreading it on a two-dimensional plane. However, Mercator type representation is the representation technique for this paper, and it is judged that visualising the heat map through VR environment is the most reasonable method in the following process. (Fig. 8 Process diagram of AI-based 3D VR heat map preference visualisation)

The two-dimensional heatmap visualisation method using biometric information is the same as this study, in which emotional data is derived and indexed using EEG and eye-tracking equipment, but the current two-dimensional heatmap representation method has limitations on the reflection of the eye-tracking heatmap on the specified image and the limited visualisation of the EEG data on the specified image. In particular, in the case of Tobii's Pro Glasses 2, which is used in the current study, the constraint of infrared lenses that cannot collect eye-tracking information on external architectural spaces when the daytime light becomes stronger, and before applying the three-dimensional VR heatmap process, the experiments were planned and conducted around one hour before and after the seasonal sunrise time. However, by applying the 3D VR heatmap process, we can not only collect eye tracking information for the early morning hours, but also conduct experiments during the daytime with strong sunlight. In addition, the application of eye-tracking heat maps using VR 360 video removes the limitation of reflecting EEG data that can only be applied to specified images. In addition, refining the existing EEG and eye-tracking data for analysis is not only time-consuming, but also highly prone to human error as researchers manually perform statistical analysis to find sections with high positive EEG and long gaze dwell time. In particular, the existing two-dimensional visualisation heat map method used by researchers does not use AI, so the time to refine and analyse the data depends on the amount of data, but it takes more than 3 hours of work, while the AI analysis of the proposed process takes less than 30 minutes from the refinement process to the analysis process, and there are fewer errors due to human error in the case of patterned analysis work. In addition, the expression of the heat map according to the specified image is also expressed through the two axes of (x,y), but in the case of the heat map based on the VR environment map, the heat map is expressed in the coordinates of the globe detective, so the heat map is expressed in the coordinates of (x,y,z). However, the z value here refers to the depth value based on the position of the VR device worn by the user, and does not mean that the heat map is represented on the three-dimensional geometry of the architectural space. In addition, since there is one more axis of coordinates, more sophisticated visualisation is possible for a three-dimensional heat map than a two-dimensional heat map (Figure 10).

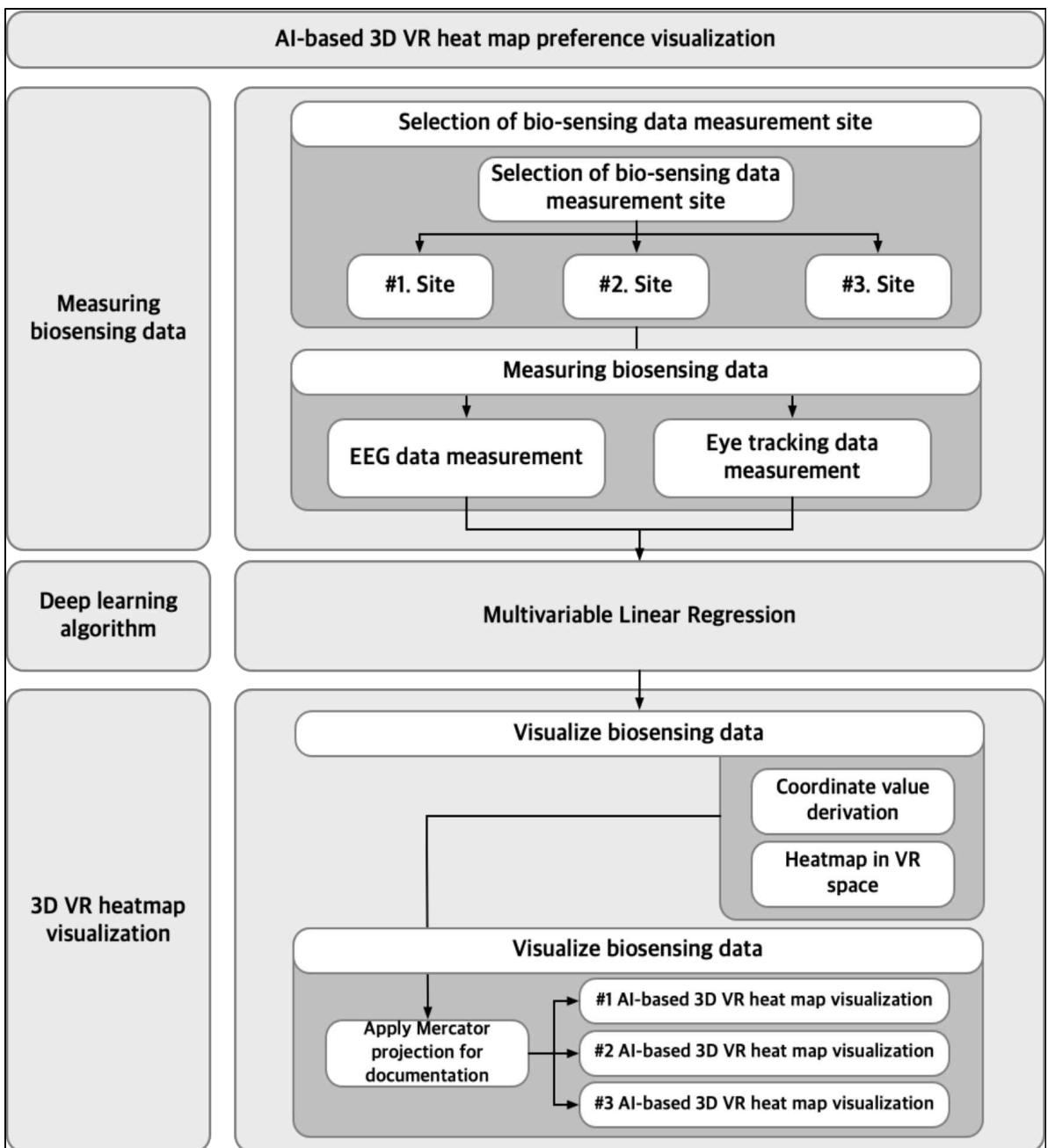

[Figure 8: AI-powered 3D VR heatmap preference visualisation process (adapted)

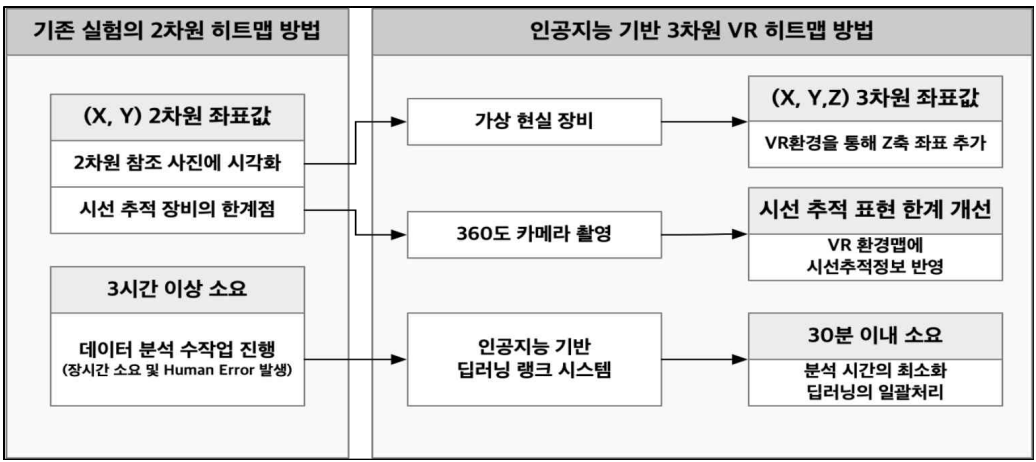

[Figure 10] Process differences between two- and three-dimensional heatmap methods in existing experiments

To solve this problem, the algorithmic process to find the coordinates with the highest overall rank, which takes into account the areas with high positive brain waves and long gaze duration, is based on Multivariable Linear Regression in deep learning, and produces the highest priority (Figure 11).

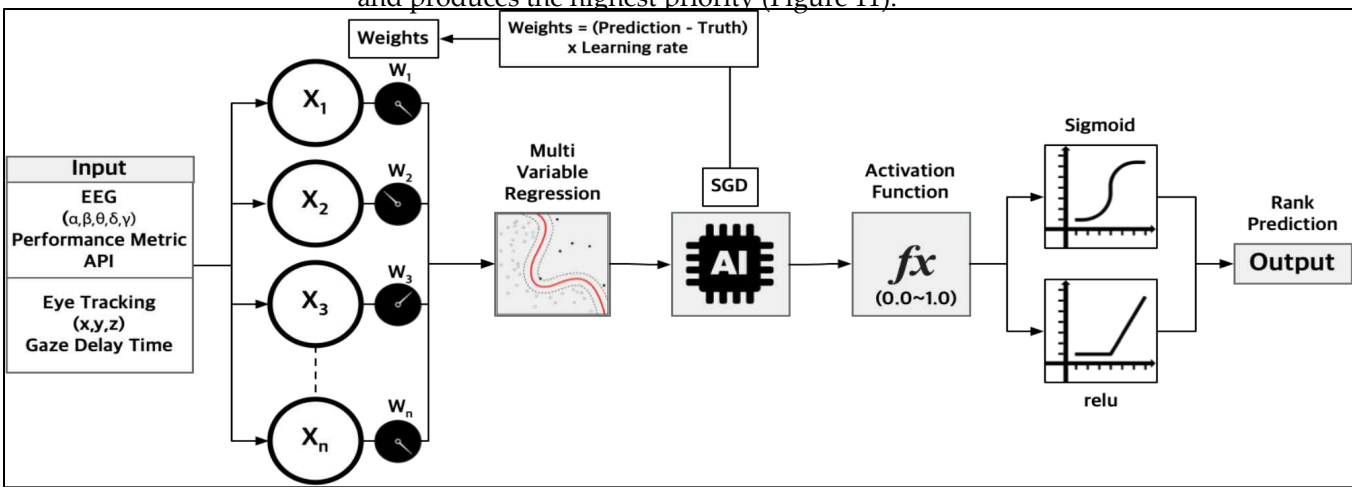

[Figure 11: Multivariable Linear Regression process diagram based on deep learning

The next process involves injecting biometric data consisting of EEG and eye-tracking information using the SGD technique and the Sigmoid activation function to find the coordinates of areas of high positive EEG, low stress, and long gaze duration. SGD is an auxiliary algorithm that finds weights for effective implementation of Multivariable Linear Regression. In particular, the weight is a deep learning algorithm that considers both emotional information from EEG as a weight and the coordinates of the gaze duration as a time axis. Sigmoid is an activation function that is compared to the relu function and is used to generate a final result in percentages between 0.0 and 1.0. Sigmoid is mainly used for ranking and classifying data, while relu is used for predicting future values based on past data patterns. In this research process, the sigmoid function was used because it is required to rank the positive emotions and eye lingering areas of the emotional information felt in the architectural space through biometric data.

## 4. Applying the biosensing-based VR heatmap process to architectural spaces

### 4.1 Collecting users' biometric data in the DDP

Apply the biosensing-based VR heatmap process to the designated architectural space of the DDP. A 360-type video recording is conducted to select the location for emotional data measurement and to measure preferences. In this study, we use the term biosensing data to incorporate data that can be monitored by sensors, such as brainwave and eye-tracking data. To select a location for the video recording of the DDP, one of the entrances to the DDP from the outside space was designated in consideration of the visitor's gaze. Then, we set up a path from the designated entrance to the outside space connected to the architectural space to create the appearance of approaching the DDP. To create the path, three videos were taken to consider the movement path, as the video coordinates are fixed for the preference heat map and the user cannot move inside the virtual space while viewing through the 360 video within the VR environment. In the proposed process, the experiments were limited to the expansion of the heatmap from a two-dimensional heatmap of a given image to a three-dimensional heatmap of a globe shape. The 360 video shooting was implemented in a three-dimensional virtual reality space using the ONE X2 camera from Insta360 (Figure 12).

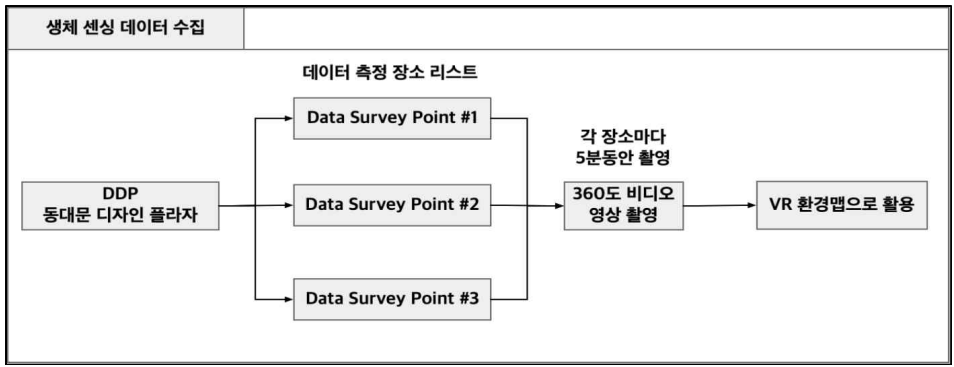

[Figure 12: Collecting biosensing data from the DDP

The data was collected by wearing an eye-tracking-based VR device that demonstrated a 360 video and an EEG device that measured biosensory data. The VR device was HTC's Vive Pro Eye, and the EEG device was Emotiv's EpocX. The Pro Eye's eye-tracking sensor collected data synchronised to the same time axis as the EEG data. EpocX is a device with a high reliability and noise filter performance index of 95.8%, and its reliability has been verified in many papers, and it has recently been used as an EEG research device in various fields. (Figure 13: The scene where I am wearing the EEG equipment and the composite image at the bottom)

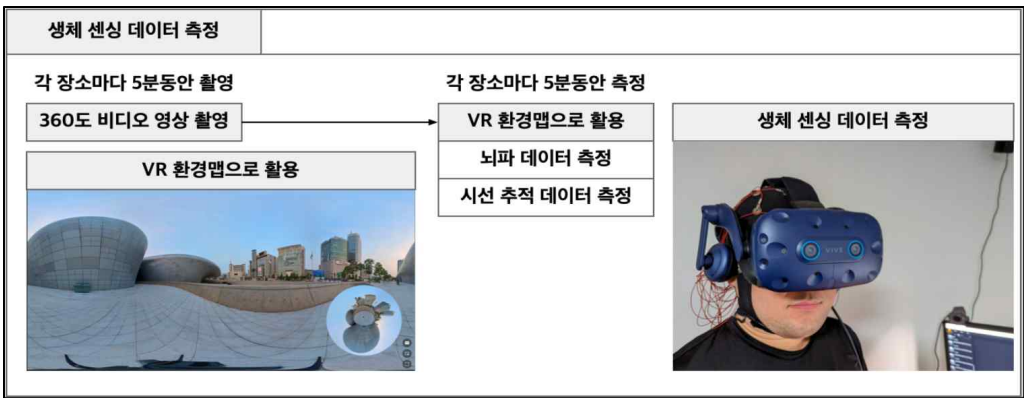

[Figure 13: The measurement process after donning the equipment for the experiment and creating the 360 video.

**4.2 Refining the user's biometric data in the DDP**

The analysis of biosensing data in this study is based on three emotional information derived from EEG data-based emotional information measurement and gaze delay time by coordinates derived from eye tracking information. Stress, Engagement, Interest, Excitement, Focus, and Relaxation are classified as emotional items converted through the Performance Metric API of the measured EEG data, and in this study, the process of ranking the time information of the points with low stress index and high interest index in the experiment in three regions was carried out, and the time information of the points with high delay time of the eye tracking information was derived, and the coordinates that intersect with the eye tracking information of the VR environment map for the highest ranking top 30 were derived to implement a heat map.

The biosensing data of this study is refined until the ranking is calculated, and the data refinement work consists of converting the raw data of EEG into emotion information through the Performance Metric API, and calculating and organising the same coordinates of delay time of the eye tracking information developed by the researcher and the coordinates of intersection with the VR environment map to form a dataset. This dataset is used as data inserted through AI model training and verification for 3D VR heatmap preference visualisation.

**4.3 Visualising users' biosensing-based VR heatmap preferences in DDP**

The preference visualisation of this study's biosensing-based VR heatmap process uses a deep learning-based ranking algorithm to derive the time information of EEG data ranked by preference and the time information of eye tracking information with high delay time. From the derived time, the intersection point of the user's gaze direction and the VR environment map is derived as a three-dimensional coordinate value, and the points in the space preferred by the architectural space experiencer are visualised as a heat map. Multivariable Linear Regression algorithm was selected and used as an algorithm to derive the ranking of situations that optimally meet the conditions, reflecting the complex conditions of low stress, high concentration, and high gaze time. Multivariable Linear Regression is a popular regression-based deep learning algorithm that can compute complex variables for each condition presented in the dataset. In addition, the sigmoid activation function and the methodology for deriving rankings due to the high frequency of use have been applied in similar situations on Kaggle, the world's largest AI prediction model and analysis competition platform founded in 2010, and the algorithm process is simple to utilise complex variables, so it was used in the process of this study. As of May 2022, the number of forums opened on the topic of Multivariable Linear Regression algorithm through Kaggle search was 208 from 2017 to May 2022, and the code was analysed and used in the process of this study. However, we believe that future research should be conducted to compare similar algorithms to select the optimal algorithm. In this study, we focused on the conversion of biosensing data from a two-dimensional heat map to a three-dimensional heat map, limiting the scope of the study to the performance analysis of AI algorithms. (Figure 15)

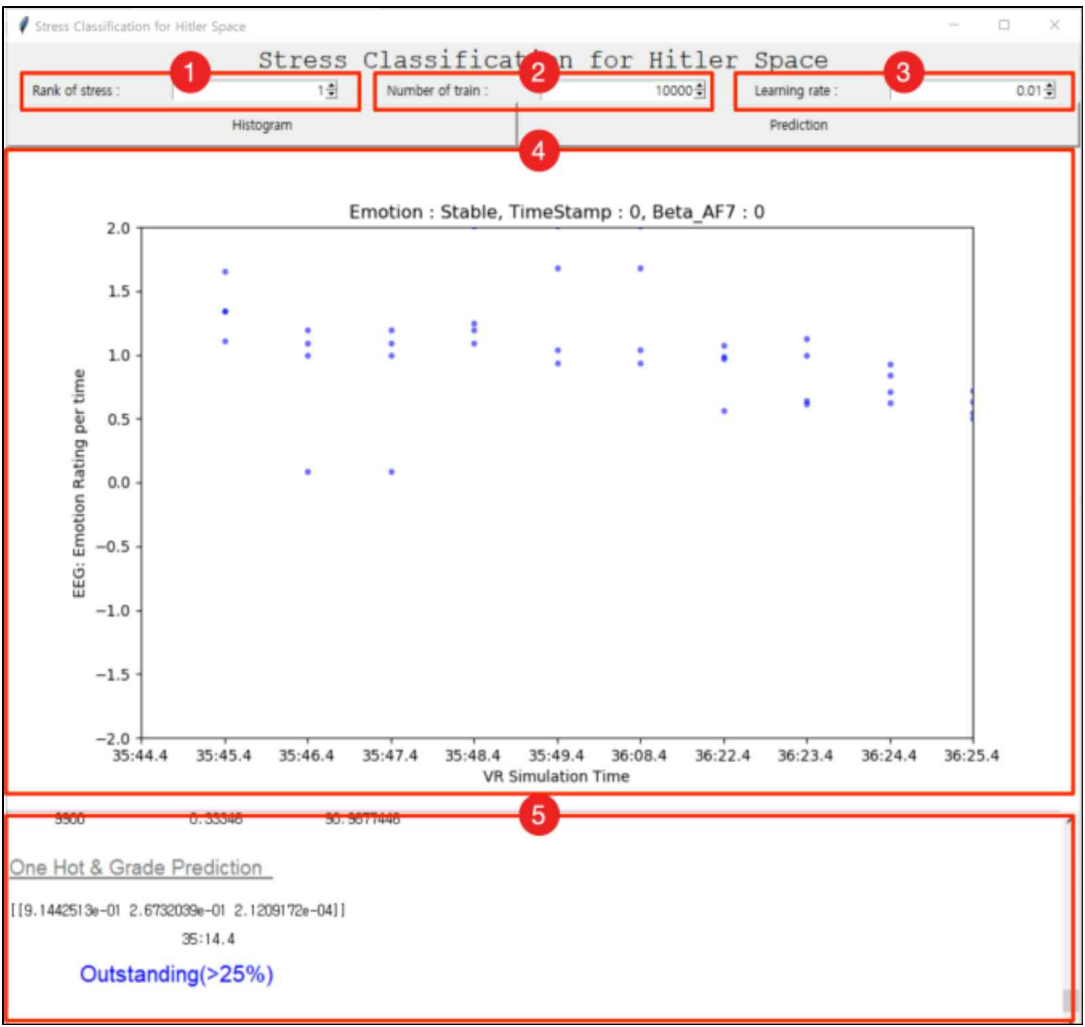

[Figure 15] Platform developed using Multivariable Linear Regression (Image at right)

The final product of this research's biosensing-based VR heatmap process is a heatmap that incorporates the user's eye tracking information and emotional data through EEG into a VR environment. VR heat map can effectively check the user's preference for architectural space through VR equipment, but in order to effectively express it through documents or this paper, the three-dimensional VR environment map is visualised in two-dimensional form in the form of pins. The heat map information expressed in the three-dimensional VR architectural space is expressed as a sphere like a globe, so the Mercator projection is used to expand the development in two dimensions. Mercator projection is one of the methods to implement the shape of a globe into a two-dimensional plane, and although the process of unfolding the map causes distortion of the 360 image, it was judged to be a universally suitable method for expressing it on a two-dimensional plane (Figure 16).

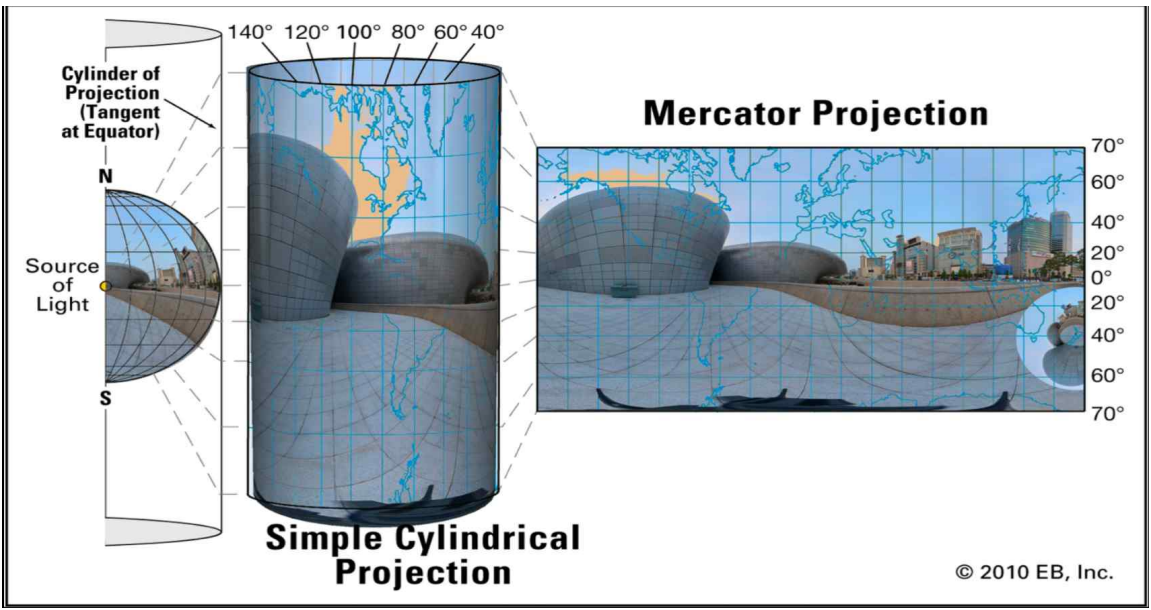

[Figure 16: Converting a three-dimensional heatmap to a two-dimensional heatmap using the Mercator method

### 4.4 Experiment and analyse the process of visualising users' biosensing-based VR heatmap preferences in DDP

### 4.4.1 Biosensing-based VR Heatmap Preference Visualisation Process Experiment Overview

To apply the two-dimensional biosensing-based VR heat mapping process, three locations were selected to capture 360-degree footage of the DDP's architectural space from the main flow of visitors to the DDP. The filmed DDP space was filmed in three locations for 10 minutes at each selected location from 6:00 to 7:00 am on 13 November 2021 (Figure 17).

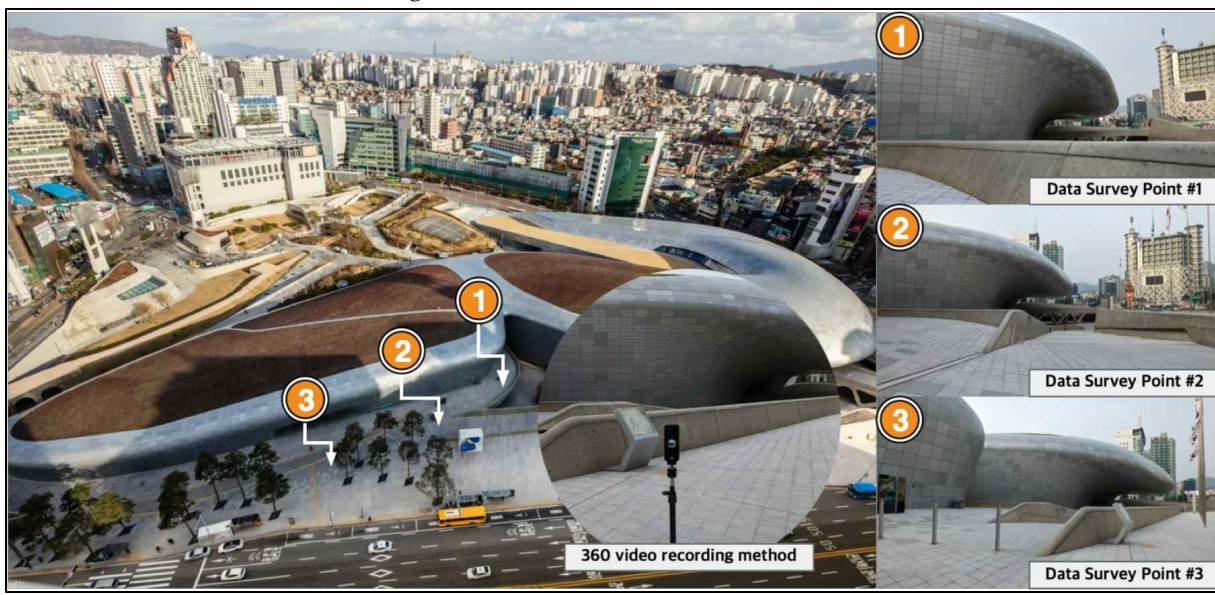

[Figure 17] Location in the DDP space where the experimental footage was taken and a photo of the experiment (for future translation of the image) https://www.kogl.or.kr/info/license.do

The biosensing data was measured in an indoor laboratory while wearing a VR environment map and EEG equipment. Three experiments were conducted on three locations where 360-degree videos were taken, with 10 minutes of measurement followed by 10 minutes of rest, for a total of three measurements. Only one experimenter, the lead author of this paper, conducted the experiments, focusing on the process of visualisation representation. Research Professor Seung-yeol Ji has been conducting EEG research in the field of architecture since 2011, and has obtained EEG data and EEG pattern profile information on individuals over a period of 10 years using only Emotiv equipment. Therefore, in the process of conducting analysis based on long-term accumulated brainwave patterns to represent the visualisation process, it was set that it is possible to analyse emotional data with high reliability on brainwaves. In this experiment, for the purpose of visualisation, we focused on the main process of heat mapping according to the planning of visualisation information and brainwave information through artificial intelligence algorithms. Therefore, rather than a study that measures and analyses brainwave data for a short period of time for various people, we conducted an experiment based on the data of one subject who has accumulated brainwave data for a long period of time to increase the reliability of the process of converting data for visualisation. In addition, we believe that the impact of applying the process in this study on the outcome of visualising individual preferences for specific buildings or spaces using EEG data and eye-tracking equipment is small. As this experiment was conducted on human subjects, it was conducted in accordance with the IRB review regulations of Hanyang University.

**4.4.2 Experimental analysis of biosensing-based VR heatmap preference visualisation process**

Through the VR experience, biosensing data is collected as EEG and eye tracking information, and the EEG information is accumulated into the PM.Scale category using the Performance Metric API. PM is a prefix for a variable that is calculated like index data when the Performance Metric API derives an index. The Stress Index, Interest Index, and eye-tracking information coordinates are then accumulated on the time axis, which the researcher wishes to explore as preferences for architectural spaces. PM.Stress.Scaled, PM.Interest.Scaled, and EyeTrace.Coordinate measured in three experiments are accumulated as shown in Table 2. Based on the accumulated data, PM.Stress.Scale is low, PM.Interest.Scale is high, and the coordinates of EyeTrae.Coordinate and EyeTrace.Coordinate are rounded to the second decimal place for each coordinate to calculate the DelayTime value to derive a value for the time the gaze stays. Using the derived values, apply an algorithm to derive a deep learning-based rank and list the top 100 rankings. Top Rank100 extracts the coordinates of the areas where PM.Interest.Scale is high and PM.Stress.Scale is low, and the coordinates of the areas where the gaze stays for a long time are the design parts of the space with high preference for architectural space. Therefore, by applying the Multivariable Regression algorithm, the preference data derived from the biosensing data analysis can be ranked and visualised on the coordinates of the 360 video on the three-dimensional VR environment map, and the experimenter's preference can be implemented through a heat map. Accordingly, the dataset based on EEG and eye tracking information was used to calculate the coordinate values for visualising the top ranked preferences after running the deep learning algorithm. The following table shows the top 15 most preferred ranks and the bottom 15 least preferred ranks (Table 2).

| Rank | Experiment 1 Building Space Coordinate Values | | Experiment 2 Building Space Coordinate Values | | Experiment 3 Building Space Coordinate Values | |
|---|---|---|---|---|---|---|
| | Preferred coordinates | Less preferred coordinates | Preferred coordinates | Less preferred coordinates | Preferred coordinates | Less preferred coordinates |
| 1 | (-35.4, -16.6, 95.0) | (-92.5, 43.5, 10.3) | (-29.0, 44, 88.3) | (-28.2, 17.3, 97.3) | (16.4, 17, 100.0) | (15.0, 16.9, 100.2) |
| 2 | (-29.3, -3.5, 98.4) | (-71.5, 22.4, 70.5) | (-26.4, 20.8, 97.2) | (-12.8, 21.9, 99.7) | (14.3, 17.8, 100.2) | (35.1, 16.4, 95.2) |
| 3 | (-57.9, 1, 85.0) | (-40.4, 20.2, 92.5) | (-56.7, 18.8, 83.7) | (-8.4, 19.2, 100.6) | (-27.9, 13.7, 97.9) | (41.2, 17, 92.6) |
| 4 | (-60.4, 22.4, 80.1) | (-17.5, 16.9, 99.8) | (-20.6, 13.2, 99.9) | (-15.8, 19, 99.8) | (-58.3, 31.2, 78.8) | (19.1, 22.5, 98.5) |
| 5 | (-42.6, 18.7, 91.7) | (-16.0, 16.2, 100.2) | (-1.4, 8.9, 102.5) | (-32.4, 20.3, 95.5) | (-55.0, 26.2, 82.7) | (-11.3, 27.2, 98.5) |
| 6 | (-29.1, 24.8, 95.4) | (20.4, 8.3, 100.4) | (-15.3, 14.2, 100.6) | (-47.6, 25.7, 87.4) | (-43.2, 23.7, 90.2) | (-57.8, 40.7, 74.7) |
| 7 | (-28.7, 8.5, 98.3) | (44.8, 7.8, 92.2) | (-11.1, 13.2, 101.4) | (-50.7, 22.6, 86.6) | (-20.4, 19, 99.0) | (-79.9, 49.8, 41.6) |
| 8 | (-33.2, 4.1, 97.1) | (67.7, 8.8, 76.8) | (-25.3, 15.2, 98.4) | (-12.7, 12.4, 101.3) | (-9.1, 18.8, 100.6) | (-60.9, 24, 79.2) |
| 9 | (-27.6, 3, 98.9) | (92.2, 3.1, 45.2) | (-46.0, 20.3, 89.7) | (2.0, 13.1, 101.9) | (-12.5, 21.9, 99.7) | (-38.7, 23.7, 92.3) |
| 10 | (-73.8, 34.7, 62.5) | (101.4, 3.7, 15.8) | (-50.8, 14.8, 88.1) | (5.7, 16.9, 101.2) | (-36.3, 16.2, 94.7) | (-16.2, 20.4, 99.5) |
| 11 | (-44.7, 26.7, 88.5) | (78.8, 6.1, 65.7) | (-49.9, 14.4, 88.7) | (-11.0, 21.7, 100.0) | (-5.0, 16.2, 101.3) | (5.0, 20.8, 100.6) |
| 12 | (-29.7, 25.4, 95.1) | (25.0, 7, 99.4) | (-30.3, 16.4, 96.9) | (-26.2, 25.3, 96.0) | (-2.9, 15.5, 101.6) | (20.0, 18.5, 99.1) |
| 13 | (-5.7, 23.8, 99.8) | (10.2, 22, 100.0) | (-31.8, 17.5, 96.2) | (-39.6, 27, 91.0) | (2.4, 15.9, 101.5) | (26.7, 20.9, 97.2) |
| 14 | (-32.6, 45.1, 86.4) | (-0.6, 17.9, 101.3) | (-4.0, 19.1, 100.9) | (-61.7, 24.7, 78.3) | (27.5, 17, 97.6) | (16.7, 22.1, 99.1) |
| 15 | (-32.2, 44.1, 87.0) | (-22.5, 15.6, 99.0) | (-1.3, 20.3, 100.9) | (-51.1, 22.1, 86.5) | (21.1, 19.4, 98.7) | (-3.8, 27.5, 98.9) |

[Table 2] Top 15 deep learning-based preference rankings using biosensing datasets

This study's biosensing-based VR heatmap preference visualisation uses the Unity3D game engine, which is easy to build a VR environment, to place the coordinate values organised by preference into a VR environment map by applying a deep learning ranking algorithm to implement a VR heatmap of the experimenter's preferred architectural space area. The three-dimensional coordinate values derived by ranking according to preference using the deep learning ranking algorithm are shown in Table 4. Then, based on the derived 3D coordinate values, we implement a heatmap that displays the preferences in the VR environment using Unity3D. The state of the heatmap is shown in the following figure (Figure 18), which shows the original VR heatmap and its development. However, the preference heatmap display that cherry-picks buildings was excluded through masking to prevent it from being displayed.

The heatmap-based preference numbers range from 100 to -100. 100 corresponds to highly preferred architectural space elements, with higher values being closer to the colour red, and -100 corresponds to less preferred architectural space elements, with lower values being closer to the colour blue. 0 is neutral in preference and corresponds to the colour Green. The following colours can be set differently according to the intention of the heatmap designer depending on the settings of the heatmap, and are limited to the representation of the results of this experiment.

The biosensing VR heatmap visualisation was visualised in a two-dimensional unfolded form to evaluate the image according to the preference of the VR heatmap implemented through emotional data visualisation (Figure 19). The visualisation area was limited to the building part, and the rest of the area represented the heat map after cropping.

In Experiment 1, the experimenter's preferred architectural areas were concentrated on the right mass building in the DDP. In particular, the preference was concentrated in the middle of the curved shape of the building, where it flows naturally, and the point where the right and left buildings meet each other. In the case of the left mass, the highest preference was measured for the top curved part of the architectural form. The less favourable architectural areas are concentrated in the lower left and upper left parts of the left mass.

Experiment 2 shows a different approach to the DDP building than in Experiment 1. The highly preferred architectural elements are similar to those in Experiment 1, but the parts of the mass that were previously highly preferred in the right-hand mass are now highly preferred in the upper right-hand part of the left-hand mass. The low-preference architectural elements are clustered in the lower left and upper left parts of the left mass, similar to Experiment 1, and in Experiment 2, the low-preference architectural elements are found in the left contact part of the left mass.

Experiment 3 is the closest to the DDP building than Experiment 2. Elements of the architectural space that were highly preferred by the experimenter were concentrated in the curves at the top of the building, while architectural elements in lower positions were less preferred. The heatmap shows that the interest index of the curved architectural elements in the mass of the building is higher when viewing the building up close.

The overall analysis shows that the experimenter tends to have a higher preference for curved shapes and parts of the building that are adjacent to the sky, and a lower preference for parts of the building that do not change the shape of the building or where the building is obscured by surrounding objects. Therefore, it is common for the experimenter to have a higher preference for curved buildings that are adjacent to the sky and curved elements at the top of the building.

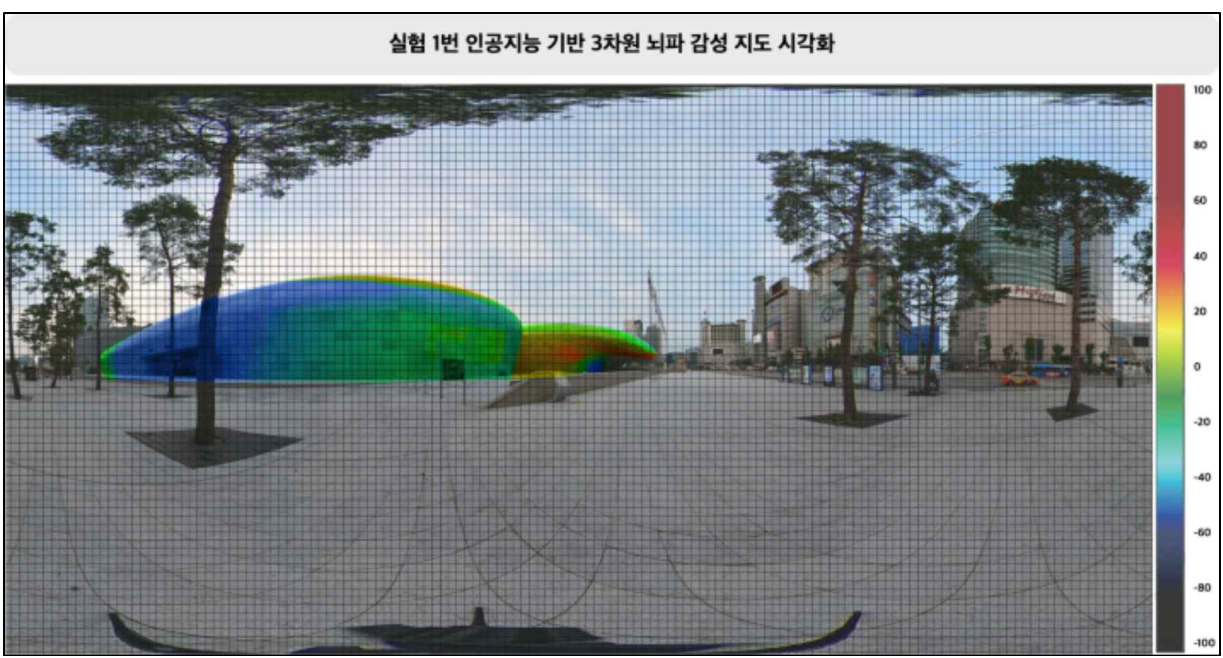

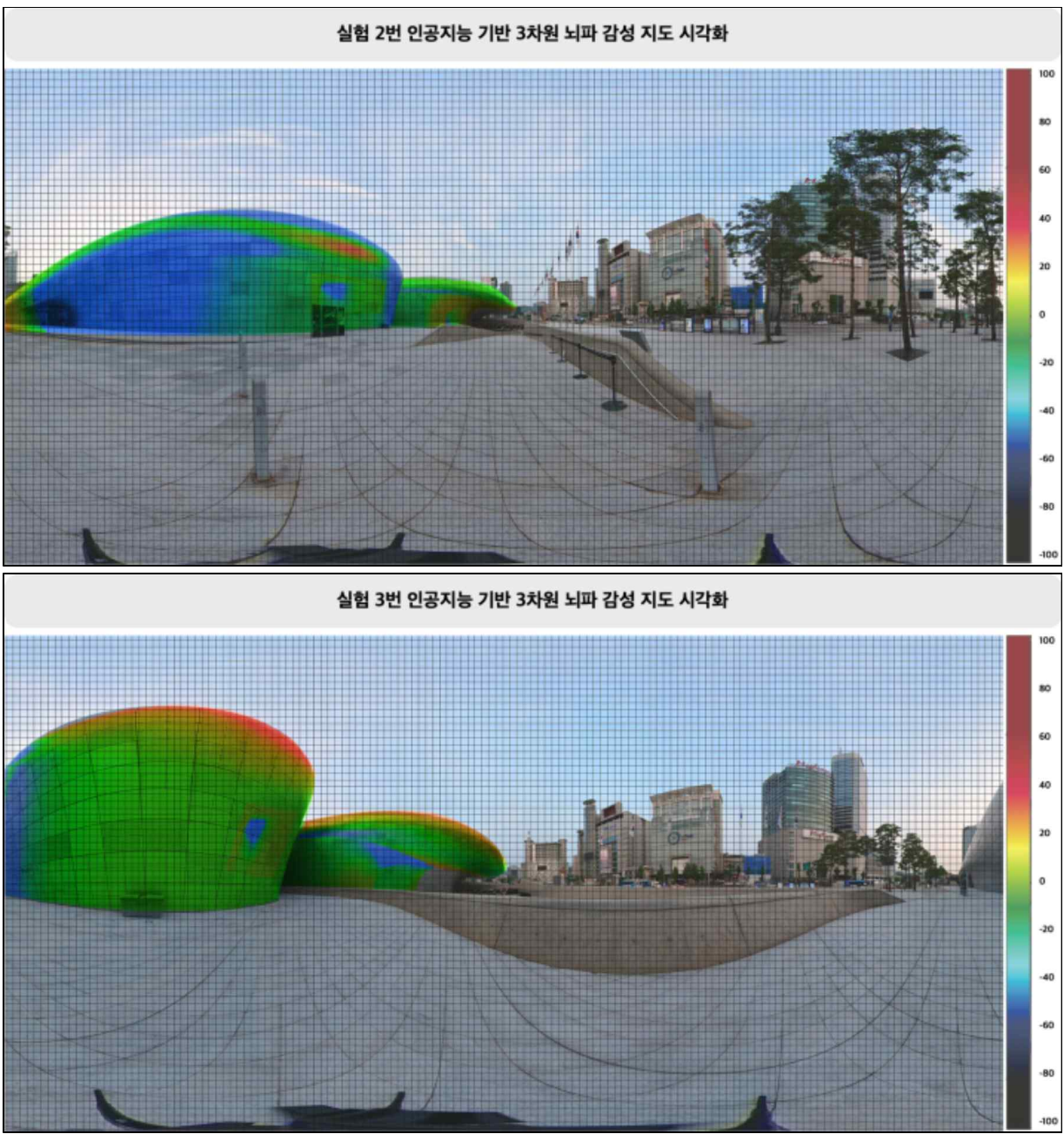

[Figure 18: Visualisation results by experiment

## 5. conclusion

In order to partially apply the method of 'collecting emotional information using maps' among the three categories of emotional maps proposed by Griffin & Mcquoid (2012) to this study, the emotional information felt while experiencing the architectural space was monitored through EEG and eye-tracking equipment, and the data was visualised as a VR heat map to identify and visualise the impact of the elements of the architectural space on humans, and to analyse the impact of the architectural form on humans through visualisation.

As a result of the analysis, the experimenter found that the curved elements at the top of the building and the parts adjacent to the sky tended to be highly preferred. On the other hand, the preference was relatively low in cases where the architectural elements were not clearly judged and in the lower part of the building. Therefore, the experimenter could confirm that the curved shape at the top of the building and the parts adjacent to the sky tended to be highly preferred only for buildings with curved elements in the DDP building. Therefore, the user's preference based on biosensing was implemented through VR heat maps.

Based on this visualisation process, the aim is to identify the emotional information of users of architectural spaces through biosensing information and create visualisation information to induce optimal architectural spaces. In particular, the process of cleaning the dataset of fragmented sensing data derived through the user's biometric data through a system, and then deriving and visualising the rank according to the planning intention of the researcher through a deep learning algorithm is an important pipeline of this research. In addition, the process of expanding the heat map to three dimensions using VR environment maps to solve the problem that the entire architectural space cannot be organised into a heat map among two-dimensional heat maps is significant. Through this, it is judged that VR through 360 video can be used to evaluate architectural spaces, reflecting trends through VR experiences of similar cases demanded by actual residents, and analysing and reflecting resident evaluation information at the time of architectural planning to provide supplementary help in design.

Therefore, through the VR heatmap visualization method of biosensing-based user preferences in architectural spaces proposed in this study, it is possible to provide residents with biosensing data using EEG equipment and eye-tracking equipment by implementing VR heatmaps based on three-dimensional design information or 360 images of similar cases created in the actual architectural design stage, and it will help architects with data for communication that can facilitate communication between residents.

This study proposed a visualisation method for architectural data that can be used as an auxiliary tool in architectural work by analysing users' emotional patterns using artificial intelligence with biosensing information using EEG equipment and eye-tracking equipment, and expressing VR heat maps through preference surveys.

In this study, a three-dimensional VR heatmap was visualised for the analysis of human preferences in architectural spaces, and a two-dimensional flat globe was presented and analysed for documentation. Preferences for architectural spaces will vary depending on the individual preference of the equipment. However, through the emotional information that humans feel in architectural spaces, we saw the possibility of analysing preferences by converting biosensing information into emotional information, and we expect that it can help architectural professionals make decisions in various planning processes. Through this, it will be possible to extract biometric data on the effects of architectural spaces on human psychology and behaviour, and to secure and use emotional data that can improve the quality of human life.

In future research, we would like to conduct a study on the performance comparison of other algorithms for effective algorithm selection, not in terms of using the Multivariable Regression algorithm as a means to rank deep learning. In addition, we would like to expand the number of research subjects to architectural experts and non-architectural experts and analyse the differences through brain waves and eye tracking, rather than focusing on the visualisation process as in this study.

**Author Contributions:**

**Funding:** This work was supported by a National Research Foundation of Korea (NRF) grant funded by the Korean government. (2021R1C1C2012502)